\documentclass[12pt]{article}
\usepackage{latexsym,amsmath,amssymb,amsbsy,graphicx}
\usepackage[cp1251]{inputenc}
\usepackage[T2A]{fontenc}
\usepackage[russian,english]{babel}
\usepackage[space]{cite} 

\begin{document}

{\bf Flux Variations in Lines of Solar EUV Radiation
Beyond Flares in Cycle 24}

\bigskip

\centerline {$E.A. ~Bruevich, G.V. ~Yakunina$}

\centerline {\it $Lomonosov ~Moscow ~State ~University, Sternberg ~Astronomical ~Institute,$}
\centerline {\it Universitetsky pr., 13, Moscow 119992, Russia}\

\centerline {\it e-mail:  {red-field@yandex.ru} }\

{\bf Abstract.} Studies in the extreme ultraviolet (EUV) and X-ray ranges of the solar spectrum are important due
to the active role of radiation of these ranges in the formation of the Earth's ionosphere. Photons of the EUV
range are completely absorbed in the upper layers of the Earth's atmosphere and induce the excitation, dissociation, and ionization of its different components and, finally, the atmospheric heating. From the archive data of the EUV Variability Experiment of the Solar Dynamics Observatory (SDO/EVE), we formed series of
diurnal values of the background fluxes radiated beyond flares in the EUV lines HeII (30.4 nm), HeI (58.4 nm),
CIII (97.7 nm), and FeXVIII (9.4 nm) in cycle 24 (from 2010 to 2017). These fluxes are compared to the corresponding values of the radio flux $ F_{10.7}$ at a wavelength of 10.7 cm and the background radiation flux $F_{0.1-0.8}$ 
in the X-ray range between 0.1 and 0.8 nm measured onboard the GOES-15 satellite of the Geostationary
Operational Environmental Satellite system. Comparative analysis has shown that the solar radiation in individual lines of the EUV range and the fluxes $ F_{10.7}$  and  $F_{0.1-0.8}$  are closely interrelated.

\bigskip

\vskip12pt
\section{INTRODUCTION}
\vskip12pt

 Solar ultraviolet (UV) radiation is a main source of
energy in the Earth's upper atmosphere. It influences
the geocosmic medium and affects the operation of
satellites and communication and navigation systems.
UV radiation changes on different time scales, from
several seconds to a year, and over an 11-year solar
cycle. Solar short-wavelength radiation forms in the
upper chromosphere, the transition zone, and the
corona, while the entire short-wavelength range
accounts for only ~9\% of the total energy. However, it
is impossible to model the state of the Earth's upper
atmosphere without observations and prognoses of the
values of the solar radiation fluxes in the UV and X-ray
spectral ranges. Variations in the short-wavelength
(SW) radiation are determined to a considerable extent
by the entire area and the evolution of structural formations in the solar atmosphere. The emergence and
development of active regions and faculae areas are
superimposed on the radiation of "the quiet Sun" and
enhance the observed variations in the UV range (Makarova et al., 1991; Lean, 1987).

Short-term variations in flares induce changes of up to
60\% in the UV range and three orders of magnitude in
the soft X-ray (SXR) range. Long-term changes in the solar
cycle cause changes to the radiation fluxes in the corresponding SW ranges from tens of percent to several
times.

The era of extra-atmospheric observations of the
Sun, i.e., observations made with instruments
onboard balloons and rockets, began in the late 1950s.
The first emission in the hard X-ray range was
detected during a flare in 1958 by Peterson and Winckler (1959). At present, extreme ultraviolet (EUV) and
soft X-ray (SXR) fluxes are continuously monitored
by instruments onboard the TIMED, SEE, SDO, and
GOES satellites (Woods et al., 2012). The GOES-15
satellite of the Geostationary Operational Environmental Satellite system is currently the main source of
data on X-ray flares on the Sun. It is expected that
such observations will be continued with a focus on
surveys of the thermosphere and ionosphere of the
Earth (Woods, 2008; Benz, 2017).

Variations in the SW radiation lead to changes in
the Earth's thermosphere and ionosphere. Photons of
UV radiation are absorbed in the upper layers of the
terrestrial atmosphere and induce ionization and dissociation of the atmospheric components, which substantially influences processes in the atmosphere and
ionosphere (e.g., Ivanov-Kholodnii and Nusinov, 1987).
The scale of changes was illustrated by Schmidtke
et al. (1981): a 30\% decrease in the total UV flux is
equivalent in value to the energy flux entering the
upper atmosphere during a strong magnetic storm.

According to Roble (1983), if the HeII line (30.4 nm)
is removed from the solar spectrum, the exospheric
temperature at the upper boundary will decrease by 88 and 129 K in the periods of the solar activity minimum
and maximum, respectively. Thus, the change in the
UV-radiation fluxes may induce a substantial response in
the thermosphere, including variations in the temperature and the optical thickness in different spectral
ranges and, consequently, in the thermospheric
energy (Kockarts, 1981).
The total electron content (TEC) in the ionosphere
is determined by the solar EUV radiation. 

The data of SW-radiation flux beyond flares substantially (several-fold) varies during the activity cycle and, naturally, depends on the general level of solar activity
(Bruevich and Yakunina, 2015).
In the paper we analyze the data from daily measurements of the fluxes, which are not related to flares,
in four UV lines - HeI, HeII, CIII, and FeXVIII at
58.4, 30.4, 97.7, and 9.4 nm, respectively. The data
were taken from the SDO/EVE observational archive.

     According to the papers by Lemen et al. (2012) and
Ivanov-Kholodnii and Nikol'skii (1969), the largest
concentration of ions emitting in the considered lines
is reached at essentially different altitudes (and under
different temperatures) in the solar atmosphere. The
lines HeII (30.4 nm; log T is equal to 4.75) and CIII (97.7 nm;
logT is equal to 4.68) form in the transition zone, while the
lines HeI (58.4 nm; log T is equal to 4.25) and FeXVIII (9.4 nm;
log T is equal to 6.7-7.0) form in the chromosphere and the
corona, respectively.

     Table 1 contains the temperatures (column 4) in
regions of the solar atmosphere (column 5) in which
the ions radiating in the specified lines are formed.
Since not all of the SDO/EVE observations completely cover cycle 24, the observational period is
shown in column 6. Column 7 contains the estimate of
variations of the mean level of the flux beyond flares
for each of the lines from the minimum to the maximum of cycle 24 (expressed in W/m2).
   Figure 1 presents variations in the UV-radiation
fluxes in the chosen spectral lines for 2010-2017. The
results of ground-based observations of the radio flux
$F_{10.7}$ are shown for comparison. For the growth phase
of cycle 24 (2010-2014), the fluxes in the HeII (30.4 nm)
and FeXVIII (9.4 nm) lines are presented. The fluxes
in the CIII (97.7 nm) and HeI (58.4 nm) lines entirely
cover cycle 24. Figure 1 and Table 1 show that the
EUV-radiation fluxes in the chosen lines change in
different ways: the largest variations are observed for
the FeXVIII (9.4 nm) line (a twofold change from the
minimum to the maximum of cycle 24), while the
smallest variations are exhibited by the most geoeffective line of the EUV spectrum, HeII (30.4 nm), (by
26\% from the minimum to the maximum of cycle 24).

\begin{figure}[tbh!]
\centerline{
\includegraphics[width=110mm]{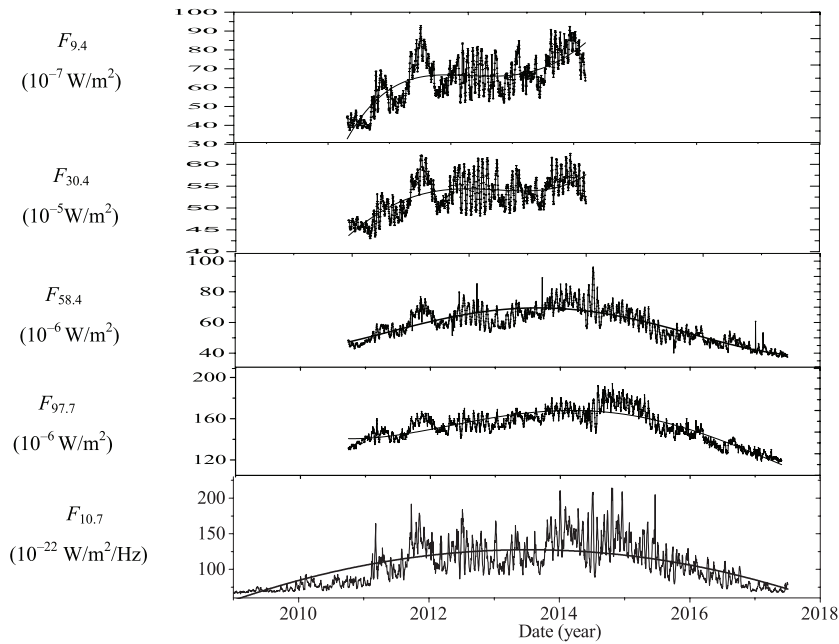}}
 \caption{Background values of the fluxes in lines of the EUV range from the SDO/EVE daily observations and the
radio flux at a wavelength of 10.7 cm in cycle 24.}
{\label{Fi:Fig1}}
\end{figure}

\begin{figure}[tbh!]
\centerline{
\includegraphics[width=110mm]{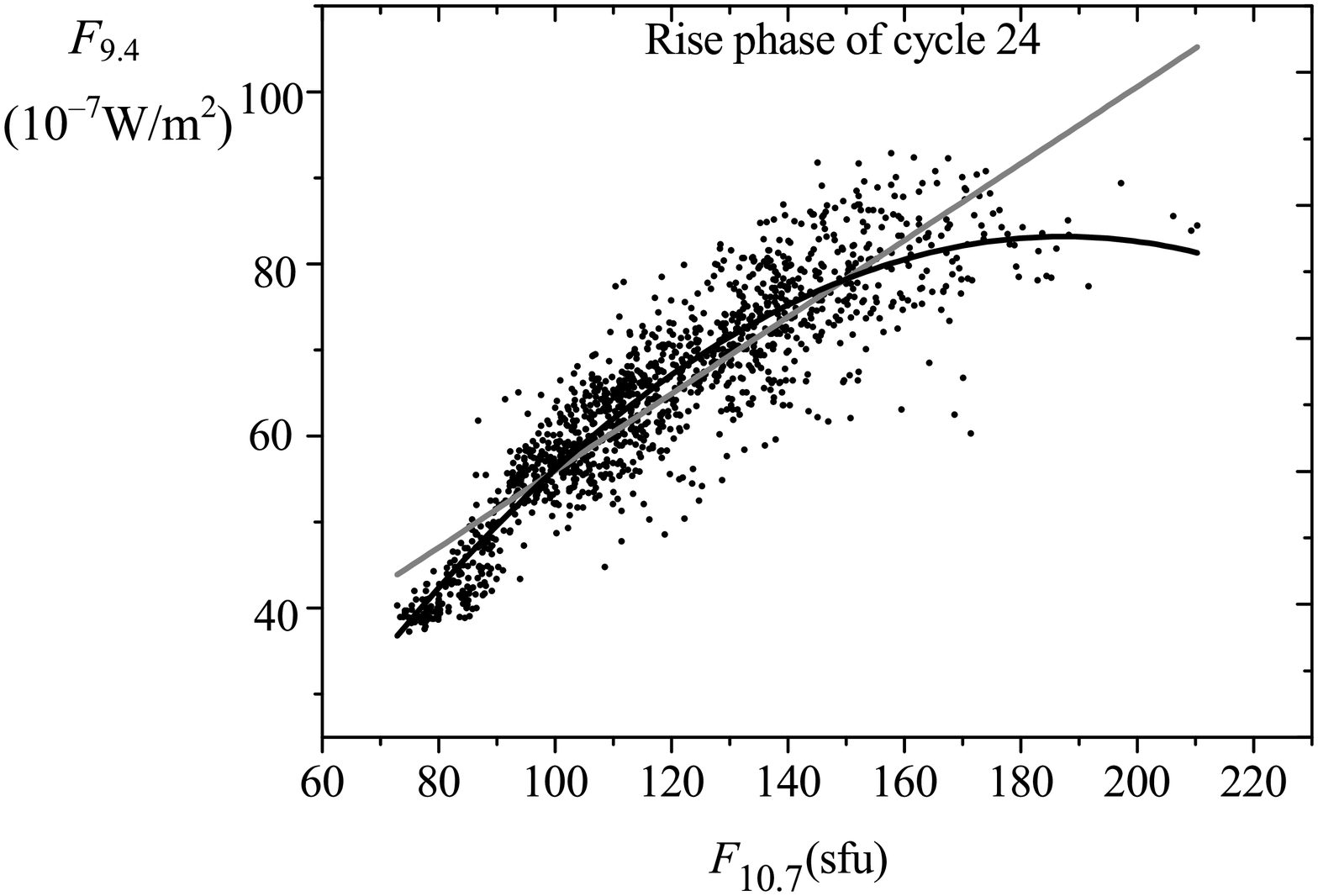}}
 \caption{Background flux in the FeXVIII line (9.4 nm) with
respect to the radio flux at the 10.7 cm wavelength in the
growth phase of cycle 24.}
{\label{Fi:Fig2}}
\end{figure}

\begin{figure}[tbh!]
\centerline{
\includegraphics[width=110mm]{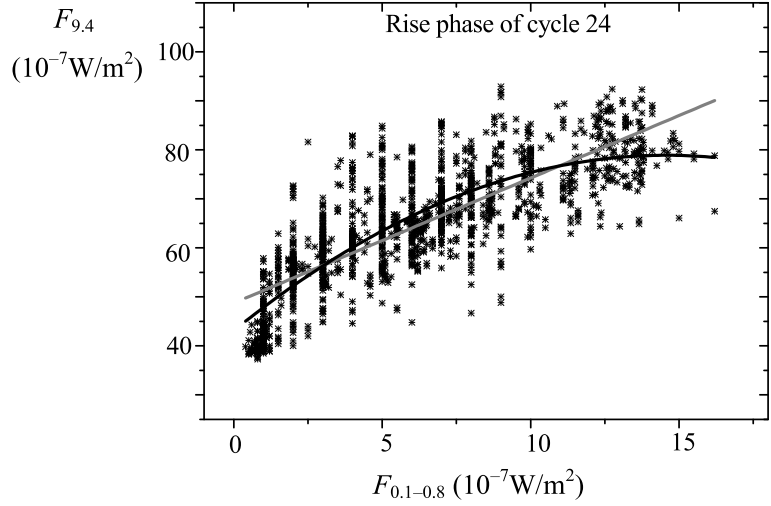}}
 \caption{Background flux in the FeXVIII line (9.4 nm) with
respect to the background X-ray flux in a range of 0.1-0.8 nm
in the growth phase of cycle 24.}
{\label{Fi:Fig3}}
\end{figure}

\begin{figure}[tbh!]
\centerline{
\includegraphics[width=110mm]{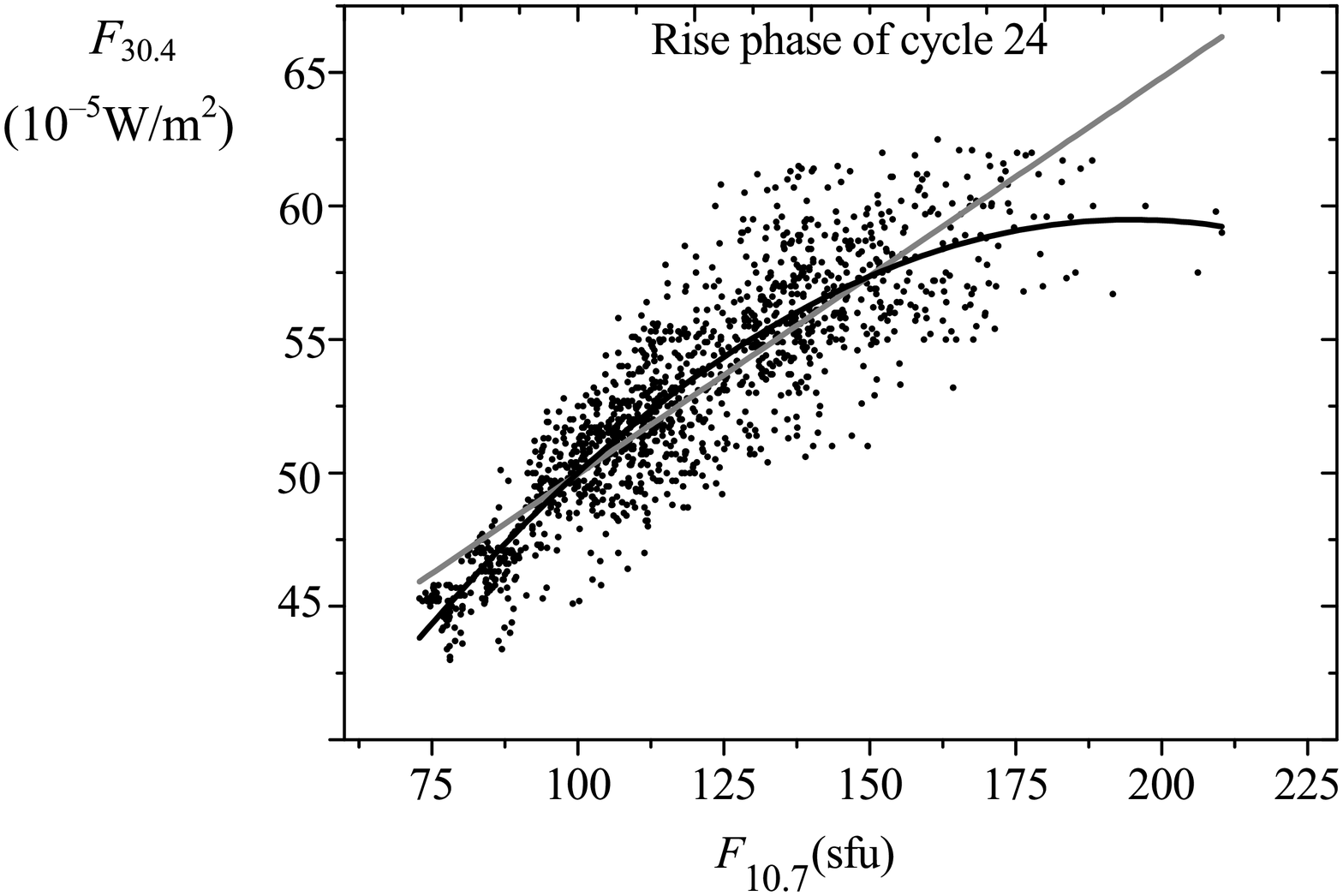}}
 \caption{Background flux in the HeII line (30.4 nm) with
respect to the radio flux at the 10.7-cm wavelength in the
growth phase of cycle 24.}
{\label{Fi:Fig4}}
\end{figure}

\begin{figure}[tbh!]
\centerline{
\includegraphics[width=110mm]{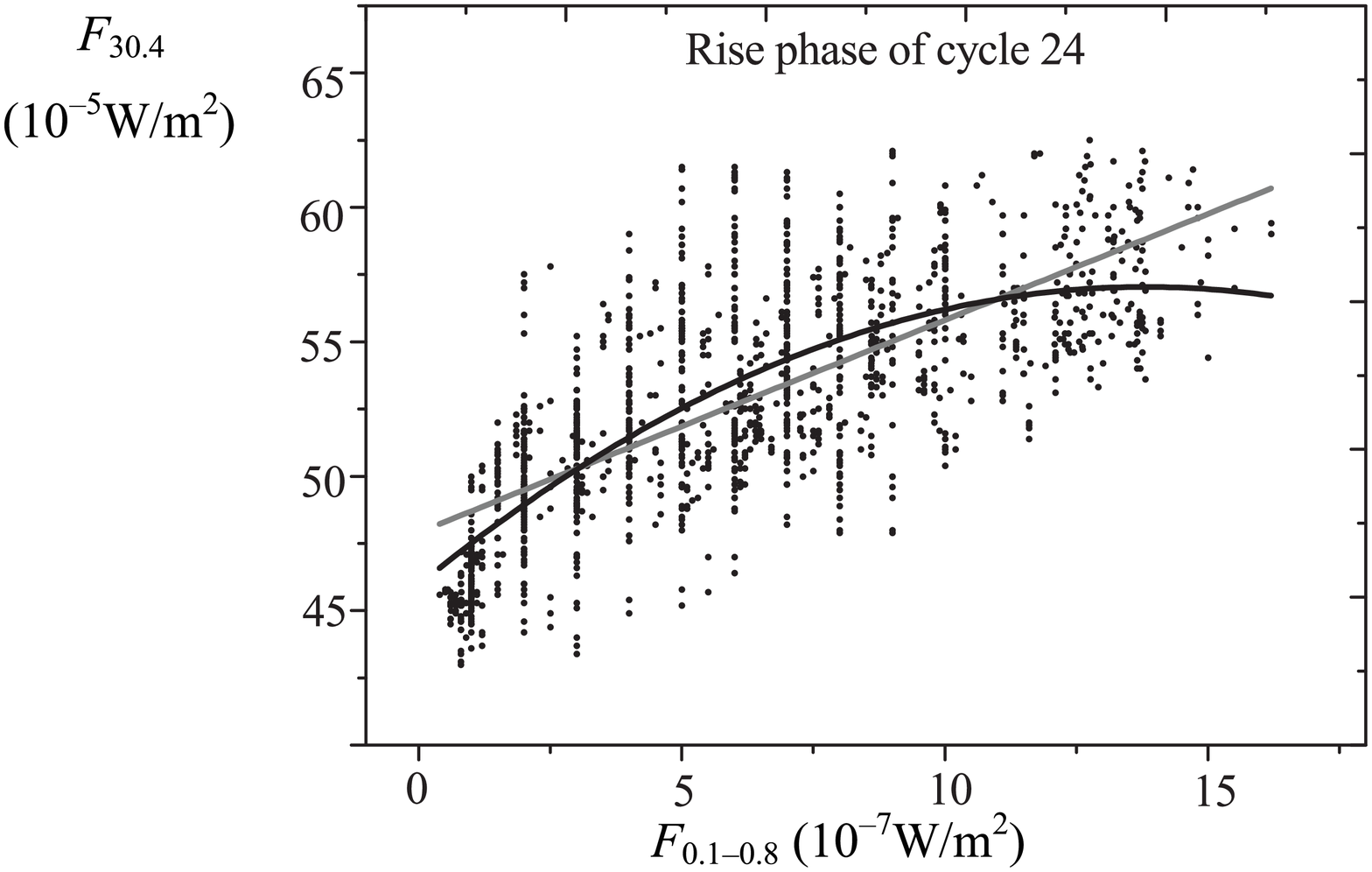}}
 \caption{Background flux in the HeII line (30.4 nm) with
respecgt to the X-ray flux in a range of 0.1-0.8 nm in the
growth phase of cycle 24.}
{\label{Fi:Fig5}}
\end{figure}

\begin{figure}[tbh!]
\centerline{
\includegraphics[width=110mm]{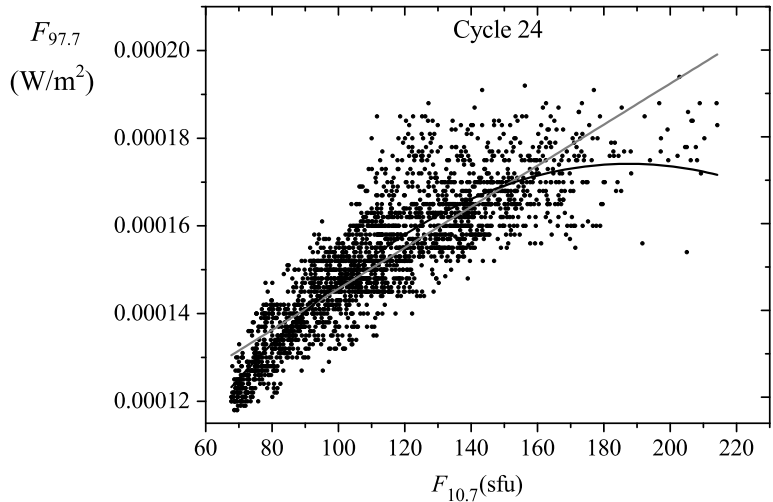}}
 \caption{Background flux in the CIII line (97.7 nm) with
respect to the radio flux at the 10.7-cm wavelength in cycle 24.}
{\label{Fi:Fig6}}
\end{figure}

\begin{figure}[tbh!]
\centerline{
\includegraphics[width=110mm]{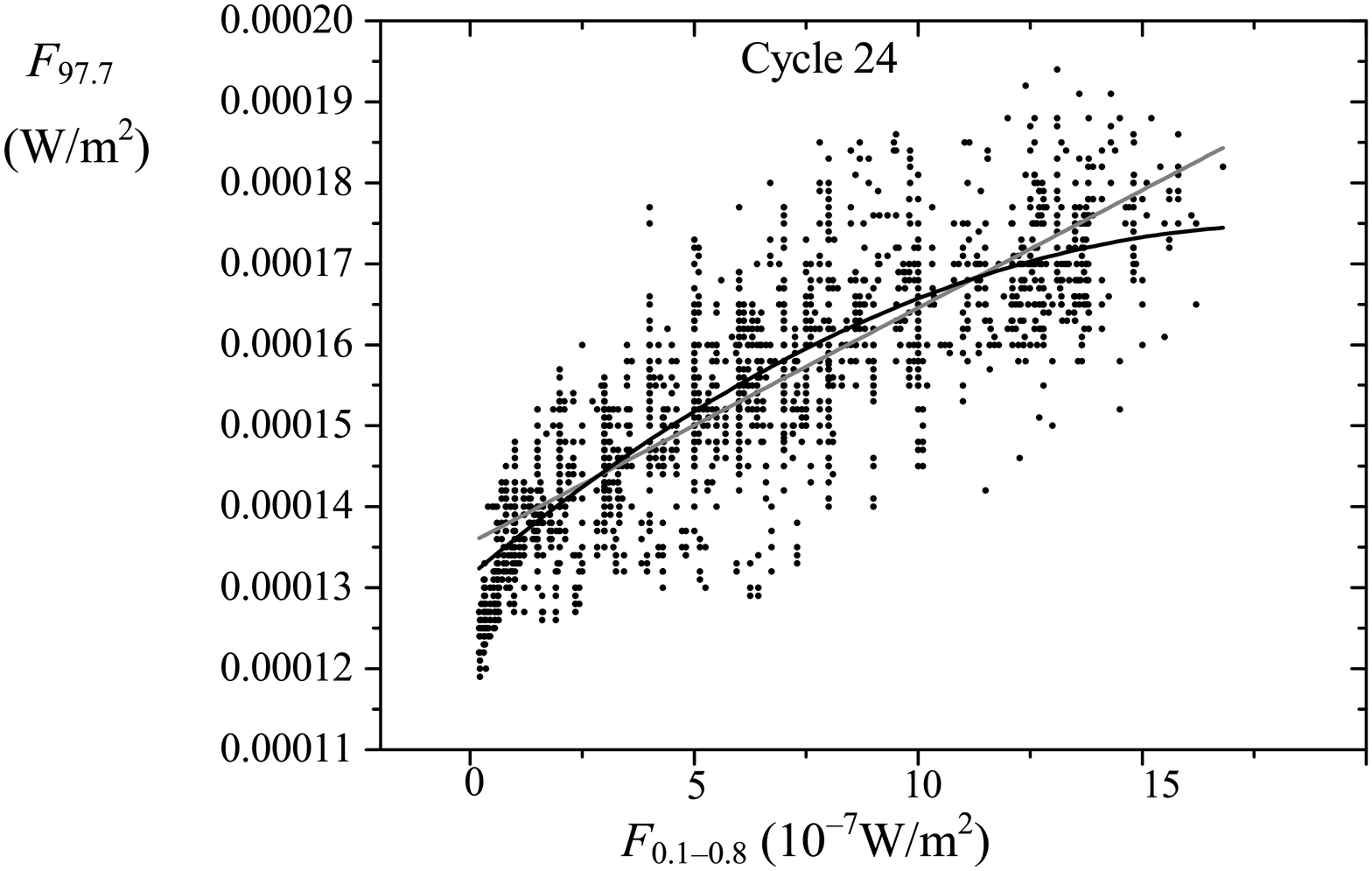}}
 \caption{Background flux in the CIII line (97.7 nm) with
respect to the background X-ray flux in a range of 0.1-0.8 nm
in cycle 24.}
{\label{Fi:Fig7}}
\end{figure}

\begin{figure}[tbh!]
\centerline{
\includegraphics[width=110mm]{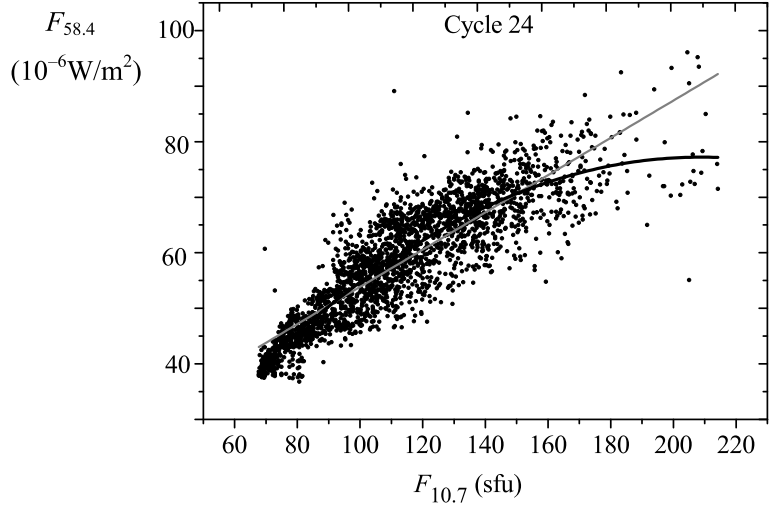}}
 \caption{Background flux in the HeI line (58.4 nm) with respect
to the radio flux at the 10.7-cm wavelength in cycle 24.}
{\label{Fi:Fig8}}
\end{figure}

\begin{figure}[tbh!]
\centerline{
\includegraphics[width=110mm]{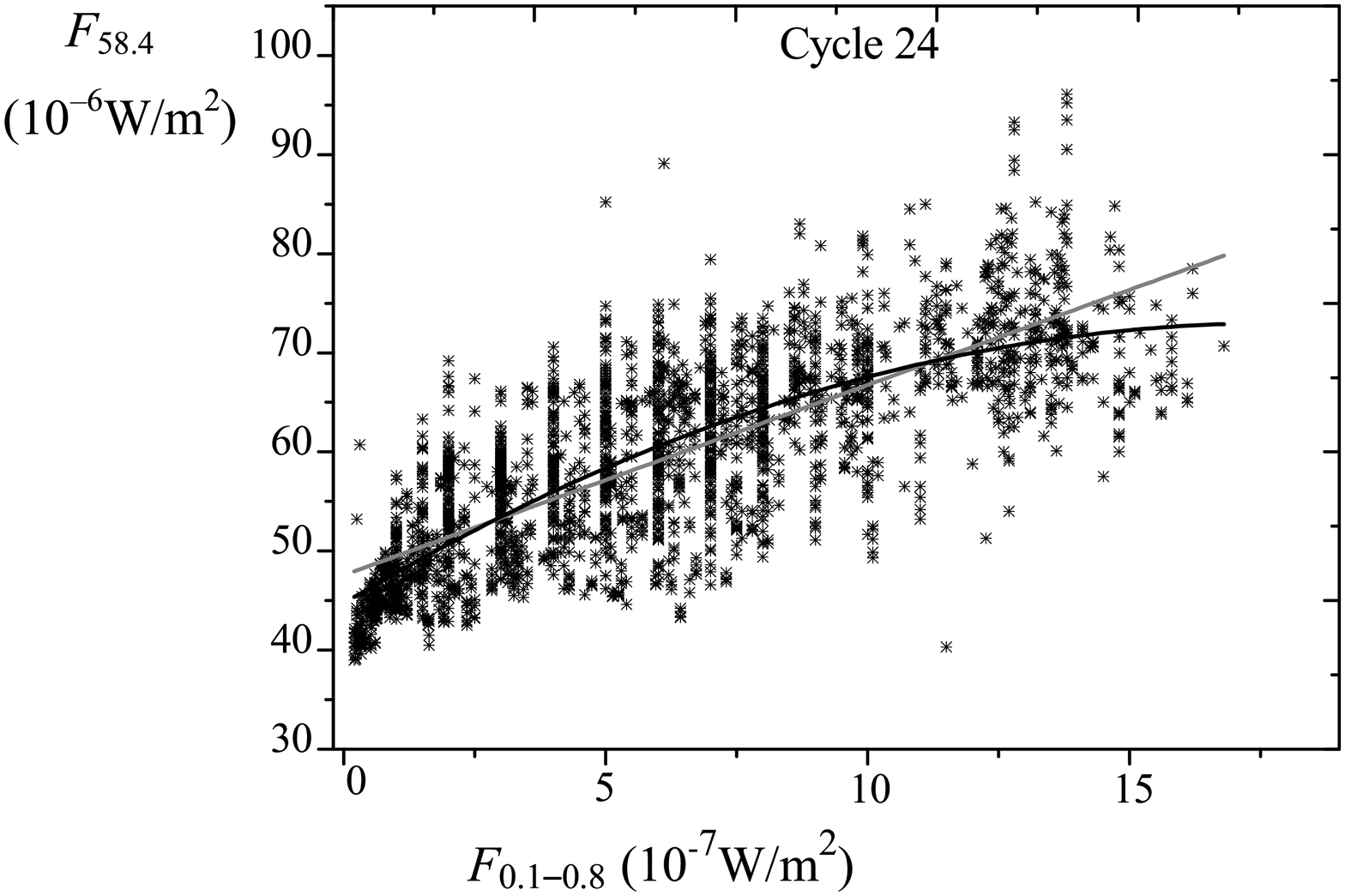}}
 \caption{Background flux in the HeI line (58.4 nm) with
respect to the background X-ray flux in a range of 0.1-0.8 nm
in cycle 24.}
{\label{Fi:Fig9}}
\end{figure}

The purpose of the paper is to study the variability
of the solar EUV-radiation fluxes and their interrelation
with the flux at a wavelength of 10.7 cm  and
background X-ray radiation in a range of 0.1-0.8 nm  in solar activity cycle 24.

\vskip12pt
\section{ FLUXES IN LINES AT THE WAVELENGTH 58.4, 30.4, 97.7,
AND 9.4 NM FROM OBSERVATIONAL DATA
OF THE SDO/EVE EXPERIMENT}
\vskip12pt

The objectives of the EUV Variability Experiment
of the Solar Dynamics Observatory (SDO/EVE) are
as follows: to study solar EUV radiation in more than
50 lines and spectral ranges, to study its variability on
different time scales to improve forecasting capabilities, and to study the flare activity effect in the EUV
range. The EVE instruments onboard the SDO satellite measure the solar EUV radiation from 0.1 to 105 nm
with a spectral resolution of 0.1 nm and a time interval
of 10 s. A spectral resolution of at least 0.1 nm is
required to distinguish the main bright emission lines,
which improves the quality of the subsequent detailed
prognosis in modeling of the Earth's ionosphere and
thermosphere (Woods et al., 2010).
Despite the success of numerical models describing
variations in the density and the temperature of the
ionosphere, the problem of verifying these models by
observations remains. The values of the solar EUV and
X-ray radiation fluxes obtained from observations turn
out to be somewhat smaller than those required to
explain the actual ion density values in the lower ionosphere ($\sim$ 110 km) (Solomon et al., 2013; Solomon,
2016).
The regions of the formation of emission lines in
the solar atmosphere depend on the temperature.
Small variations in the UV and X-ray radiation connected with appearing and disappearing groups of
spots in active regions and with variations of activity in
the solar cycle and large flares cause noticeable
changes in the UV and X-ray indices of activity.

Simon (1981) showed that, according to the measurements in cycle 20, the amplitudes of variations in the
UV radiation strongly depend on the wavelength: they
decrease increasing wavelength (a twofold difference,
less than 1\% for intervals of 135-175 and 330-340 nm,
respectively). Rottman (1988) reports an analogous
result for solar activity cycle 21: there is a twofold
change in radiation variations from the maximum to
the minimum in the range 121.6-150 nm and a
decrease to 20\% for$\lambda$  = 190 nm.

The GOES, SDO, ACE, SOHO, Proba 2, and STEREO
satellites, combined with ground-based data such as
the $ F_{10.7}$  index, are used for real-time TEC forecasting
(Hinrichs et al., 2016). Didkovsky and Wieman (2014)
and Jee et al. (2014) analyzed the TEC response to the
EUV-radiation variations in the 30.4-nm line in a
period of 1995-2013. It turned out that the EUV-radiation fluxes exhibit considerable relative variations in
activity, which may reach 20\% in the cycle.

\begin{table}
\caption{Data on the background fluxes in lines during observations of the SDO/EVE experiment in the line-forming regions}
\begin{center}
\begin{tabular}{clclclclclclcl}

\hline

N&Ion& $\lambda$, nm&log$T_i$&Line formation&Observations&Flux range, $W/m^2$ \\ 

\hline
1 &2 &3 &  4&5& 6&7 \\
\hline
1 & FeXVIII &9.4  &  7.0 & Corona&09.2010-06.2014&4E-6 - 8E-6 \\
2 &HeII& 30.4 &  7.0 & Trans. zone &09.2010-06.2014&4.6E-5 - 5.8E-5 \\
3& HeI& 58.4& 4.25&Chromosp.&09.2010-06.2017&4.65E-5 - 7.6E-5\\
4& CIII& 97.7& 4.68& Trans. zone&09.2010-06.2017&1.35E-4 -1.8E-4\\
5&$F_{10.7}$&10.7 cm& 6.0-6.5& Corona&01.2008-06.2017&67 sfu - 260 sfu\\

\hline

\end{tabular}
\end{center}
\end{table}

Note than in Table 1  the value of 1 sfu is equal to $10^{-22} W/m^2/Hz $ ,
The temperatures (column 4) are taken from the Lemen et al. (2012) and Ivanov-Kholodnii and Nikiol'skii (1969).

\vskip12pt
\section{CONNECTION OF THE FLUXES IN EUV LINES WITH THE FLUXES AT A WAVELENGTH OF 10.7 CM AND AN INTERVAL OF 0.1-0.8 NM}
\vskip12pt

     The high correlation degree of $F_{10.7}$ with all of the
main activity indices suggests that the indices strongly
depend on the plasma parameters where these fluxes
are formed, while the regions of their formation are
spatially close.  $F_{10.7}$ monitoring is a useful instrument
to predict variations in the solar coronal UV radiation,
which changes by an order of magnitude in dependence on the number and the brightness of the solar
active regions. UV fluxes play an important role in the
heating of the Earth's atmosphere and, consequently, in forming the Earth's climate. Since the total flux
$F_{10.7}$ correlates rather well with the integrated fluxes in
the UV and EUV ranges of the solar spectrum, it can
also be used as a basic index to predict fluxes in these
intervals of the solar spectrum. The radio$F_{10.7}$ from
the entire solar disk can be divided into three components (according to the characteristic time scales):
(1) events connected with flare activity of less than
one hour; (2) slow intensity variations lasting several
years that are connected with the evolution of active
regions in a solar cycle (the S component); (3) the
lowest $F_{10.7}$ level, below which the intensity never
drops, i.e., the so-called quiet-Sun level. According to
observations, there is a close correlation between the S
component of the radio emission at 10.7 cm with
fluxes in the UV lines (Tapping, 2013). The flux at a
wavelength of 10.7 cm grows with increasing temperature, material density, and magnetic fields, which
makes it a good indicator of the general level of solar
activity (Bruevich and Yakunina, 2015).
In the paper, we use data on daily SDO/EVE
observations of the background fluxes in the EUV
range and data on daily observations of $F_{0.1-0.8}$ performed beyond flares in cycle 24, which we
obtained from the archive of the GOES-15 satellite data.
 For these
observations, we built the linear and quadratic dependences of the EUV fluxes in different lines on both the
radio$F_{10.7}$  and the soft X-ray $F_{0.1-0.8}$  for the growth
phase of cycle 24 (2010-2014) and the entire cycle 24
(2010-2017). These dependences are presented in
Figs. 2-9; the linear and quadratic regressions are
shown with gray and black solid lines, respectively.
    Table 2 contains the quadratic regression coefficients (A, B1, and B2) for the dependences of the Fline
fluxes in four analyzed lines of the EUV range on the
$F_{10.7}$ value and the background $F_{0.1-0.8}$  in cycle 24. They
correspond to the following equations:

$$ F_{Line} = ~A + B1 \cdot F_{10.7}+ B2  \cdot F_{10.7}^2 $$

and

$$ F_{Line} = ~A + B1 \cdot F_{.1-0.8}+ B2  \cdot F_{0.1-0.8}^2 $$

where A, B1, and B2 are the quadratic regression coefficients,$ F_{Line}$ are the fluxes in analyzed lines, $F_{10.7}$ is the
radio flux at a wavelength of 10.7 cm, and $F_{0.1-0.8}$  is the
X-ray flux beyond flares in a range of 0.1-0.8 nm.
The values of the residual sum of squares (RSS,
which is a measure of the difference between the
observational data and the regression-line model)
were estimated for the linear and quadratic regressions. The small RSS value indicates a close fit with
the data by the model. In our case, the RSS values are
smallest when the dependence of the activity index on
the  $F_{10.7}$ value is described by second order polynomials.

Table 2 shows that there is a strong correlation
between the UV fluxes in the analyzed lines and the $F_{10.7}$.
The corresponding correlation with the$F_{0.1-0.8}$ turned
out to be much weaker.

\begin{table}
\caption{ Quadratic regression coefficients for the dependences of the fluxes in four lines of the EUV range on $F_{10.7}$  and the
background flux $F_{0.1-0.8}$}
\begin{center}
\begin{tabular}{lclclclclclcl}

\hline

$F_{Line} - F_{10.7}, F_{0.1-0.8} $& A& B1 & B2 & $\sigma A$ & $\sigma B1$& $\sigma B2$ \\ 

\hline
1 &2 &3 &  4&5& 6&7 \\
\hline
$F_{30.4}-F_{10.7}$& 	1.95E-4 &	4.11E-6 &	-1.05E-8 &	1.03E-5&	1.69E-7 &	6.75E-10\\
$F_{30.4}-F_{0.1-0.8}$ &4.59E-4 &	1.60E2& 	-5.81E7 &	2.24E-6 &	7.37 &	3.1E3\\
$F_{9.4}-F_{10.7}$&-4.12E-6&  1.33E-7& 	-3.54E-10 &	2.66E-7 &	4.36E-9& 	1.74E-11\\
$F_{9.4}-F_{0.1-0.8}$& 4.31E-6& 	4.92 &-1.69E6 &	5.72E-8&	0.19& 1.29E5\\
$F_{97.7}-F_{10.7}$&	4.94 E-5& 	1.33E-6 &	-3.54E-9 &	2.11E-6 &	3.56E-8 &	1.45E-10\\
$F_{97.7}-F_{0.1-0.8}$&4.94E-5 &	1.33E1 &	-3.55E6 &	2.11E-6 &	3.57E-1 &	1.45E4\\
$F_{58.4}-F_{10.7}$& 	-5.95E-6&	7.97E-7& 	-1.91E-9 &	1.37E-6 &	2.33E-8 &	9.53E-11\\
$F_{58.4}-F_{0.1-0.8}$&4.47E-5 &	3.17E1 &	-8.9E6 &	0.33E-6 &	1.11 & 7.52E4 \\

\hline

\end{tabular}
\end{center}
\end{table}

\vskip12pt
\section{Conclusions}
\vskip12pt

Solar EUV-radiation is the main source of heating
and ionization of the upper layers of the terrestrial
atmosphere. It is completely absorbed by the Earth's
atmosphere and governs the main parameters of the
upper atmosphere. It is important to study the EUV-radiation variations 
(diurnal and long-term in an 11-year cycle), because they contain information on the
solar chromosphere and corona and the processes
during solar flares.
Our analysis yielded the following results.
From the SDO/EVE observational data archive,
we formed four data sets containing the diurnal values
of the background fluxes in the lines HeI (58.4 nm),
HeII (30.4 nm), CIII (97.7 nm), and FeXVIII (9.4 nm)
during the 2010-2017, 2010-2014, 2010-2017, and
2010-2014 periods, respectively. The variations of
these fluxes in the considered lines depend on the
wavelength to different extent; they are presented in
Fig. 1 and Table 1.
Regression analysis showed that the radiation in
the considered lines of the EUV range was closely connected with the radio $F_{10.7}$ and the flux in the soft
X-ray range. We used second-order regression equations, since the quadratic regression yields a RSS value
substantially smaller than that produced by the linear
regression. Table 2 shows the results of the analysis.
From the results of regression analysis (Table 2),
the values of the fluxes in the considered lines can be
retrieved with the use of the data on  $F_{10.7}$  and $F_{0.1-0.8}$ 
available in real-time mode.

\vskip12pt
{\bf{REFERENCES}}
\vskip12pt

-Benz, A.O., Flare observations, Liv. Rev. Sol. Phys.,2017,
vol.14, id 2.

-Bruevich, E.A. and Yakunina, G.V., The cyclic activity of
the sun from observations of the activity indices at different time scales, Moscow Univ. Phys. Bull., 2015, vol. 70, no. 4, pp. 282-290.

-Didkovsky, L. and Wieman, S., Ionospheric total electron
contents (TECs) as indicators of solar EUV changes
during the last two solar minima, J. Geophys. Res.:
Space, 2014, vol. 119, pp. 4175-4184.

-Hinrichs, J., Bothmer, V., Mrotzek, N., et al., Impacts of
space weather effects on the ionospheric vertical
Total Electron Content, in Proc. EGU General
Assembly, 17-22 April 2016, Vienna, Austria, 2016,
id EPSC2016-7375.

-Ivanov-Kholodnyi, G.S. and Nikol'skii, G.M., Solntse i
ionosfera (The Sun and the Ionosphere), Moscow:
Nauka, 1969.

-Ivanov-Kholodnyi G.S. and Nusinov, A.A., Shortwave
solar radiation and its influence on the upper atmosphere and ionosphere, Itogi Nauki Tekh., Ser. Issled.
Kosm. Prostranstva, 1987, vol. 26, pp. 80-154.

-Jee, G., Lee, H., and Solomon, S.C., Global ionospheric
total electron contents (TECs) during the last two solar
minimum periods, J. Geophys. Res.: Space, 2014,
vol. 119, pp. 2090-2100.

-Kockarts, G., Effects of solar variations on the upper atmosphere, Sol. Phys., 1981, vol. 74, pp. 295-320.

Lean, J., Solar ultraviolet irradiance variations. A review,
J. Geophys. Res., 1987, vol. 92, pp. 839-868.

-Lemen, J.R., Title, A.M., Akin, D.J., et al., The Atmospheric Imaging Assembly (AIA) on The Solar Dynamics Observatory (SDO), Sol. Phys., 2012, vol. 275,
pp. 17-40.

-Makarova, E.A., Kharitonov, A.V., and Kazachevskaya, T.V.,
Potok solnechnogo izlucheniya (Solar Radiation Flux),
Moscow: Nauka, 1991.

-Peterson, L.E. and Winckler, J.R., Gamma-ray burst from
a solar flare, J. Geophys. Res., 1959, vol. 64, pp. 697-
708.

-Roble, R.G., Dynamics of the earth's thermosphere, Rev.
Geophys. Space Phys., 1983, vol. 21, pp. 217-233.
Rottman, G.J., Observations of solar UV and EUV variability, Adv. Space Res., 1988, vol. 8, pp. 53–66.

-Schmidtke, G., Bursken, N., and Sunder, G., Variability of
solar EUV fluxes and exospheric temperatures, J. Geophys. Res., 1981, vol. 49, pp. 146-148.
Simon, P.C., Solar irradiance between 120 and 400 nm and
its variations, Sol. Phys., 1981, vol. 74, pp. 273-291.

-Solomon, S.C., Solar soft X-rays and the ionosphere Eregion problem, in American Geophysical Union. Fall General Assembly, 2016, id SH11D-02.

-Solomon, S.C., Qian, L., and Burns, A.G., The anomalous
ionosphere between solar cycles 23 and 24, J. Geophys.
Res.: Space, 2013, vol. 118, pp. 6524-6535.

-Tapping, K.E., The 10.7 cm solar radio flux (F10.7), Space
Weather, 2013, vol. 11, pp. 394-406. 

-Woods, T.N., Recent advances in observations and modeling of the solar ultraviolet and X-ray spectral irradiance, Adv. Space Res., 2008, vol. 42, pp. 895-902.

-Woods, T., Eparvier, F., Hock, R., et al., First light results
from the SDO Extreme Ultraviolet Variability Experiment (EVE), in 38th COSPAR Scientific Assembly, Bremen, Germany, 2010, pp. 8-11.

-Woods, T.N., Eparvier, F.G., Hock, R., et al., Extreme
Ultraviolet Variability Experiment (EVE) on the Solar
Dynamics Observatory (SDO): Overview of science
objectives, instrument design, data products, and
model developments, Sol. Phys., 2012, vol. 275,
pp. 115-143.

\end{document}